\documentclass[prl,twocolumn,showpacs,preprintnumbers,amsmath,amssymb,superscriptaddress]{revtex4}

\usepackage{graphicx}
\usepackage{dcolumn}
\usepackage{bm}

\begin{document}
\title{Nano-scale brushes: How to build a smart surface coating}

\author{Holger Merlitz}
\email{merlitz@gmx.de}
\affiliation{Department of Physics and ITPA, Xiamen University,
Xiamen 361005, P.R.\ China}
\affiliation{Leibniz-Institut f\"ur Polymerforschung Dresden, 01069 Dresden,
Germany}
\author{Gui-Li He}
\affiliation{Institut f\"ur Theoretische Physik II,
  Heinrich-Heine-Universit\"at D\"usseldorf, 40225 D\"usseldorf, Germany
}
\author{Chen-Xu Wu}
\affiliation{Department of Physics and ITPA, Xiamen University,
Xiamen 361005, P.R.\ China}
\author{Jens-Uwe Sommer}
\affiliation{Leibniz-Institut f\"ur Polymerforschung Dresden, 01069 Dresden,
Germany}

\date{\today}

\begin{abstract}
Via computer simulations, we demonstrate how a densely
grafted layer of polymers, a {\it brush}, could be turned
into an efficient switch through chemical modification of some
of its end-monomers. In this way, a surface coating with
reversibly switchable properties can be constructed. We analyze 
the fundamental physical principle behind its function, a recently 
discovered surface instability, and demonstrate that the combination
of a high grafting density, an inflated end-group size and
a high degree of monodispersity are conditions for an optimal   
functionality of the switch.
\end{abstract}

\pacs{64.70.km, 64.70.Nd, 64.75.St}

\maketitle

The creation of functional materials with extreme properties
has become a key issue of modern nanotechnology. Countless potential 
fields of application have been spotted, including adhesives which
stick on any surface~\cite{creton_MRS03}, superhydrophobic 
coatings~\cite{liu_Nanotech06} or anti-friction 
coatings~\cite{li_MST07}, to cite just a few of them. Functional
surfaces which are also able to adapt to their environment 
through a reversible switching mechanism are leading another step 
further. A polymer brush, i.e.\ a coating made of polymers which
are densely grafted with one end onto a substrate, is regarded a good candidate 
for such a high-tech coating. Current approaches suggest the use
of mixed brushes, one component of which would consist of hydrophilic 
and the second of hydrophobic polymers. A change of the 
solvent would then flip up the respective compatible layer
while hiding the incompatible one. The reversible switching of 
such a mixed polymer brush was first observed in experiment by  
Sidorenko~\cite{Sidorenko_LAN99} and analyzed in detail by
Minko~\cite{minkoMM_RAPID_COMM01} and 
Lemieux~\cite{lemieux_MM03}. Motornov et al.\ have
recently applied this technology to modify the surface of colloids 
in order to create stimuli-responsive (``smart'') 
nanoparticles~\cite{motornov_AdvFunctMater07}.

This article is going to present another approach and offering,
through off-lattice computer simulations,
a proof of principle for the reversible switching of homopolymer
brushes (i.e.\ brushes consisting of only one polymer species),
grafted at high density and featuring a modified end-monomer.
Such high density brushes have recently been successfully 
created in the laboratory of Devaux et al.\ using a specially
modified grafted-from technique~\cite{devaux_MM05}.
A systematic theoretical investigation of these high density
brushes is presently a matter of intense research 
activities~\cite{biesheuvel_MM08,coluzza_PRL08,merlitz_MM08}.

Perhaps the most striking feature of a high density 
brush is its sharp drop of monomer density near the
surface, fundamentally different from the parabolic 
density profiles commonly found in low density brushes~\cite{semenov_JETP85,milner_MM88}. This
box-like profile has been predicted by
modified SCF theories~\cite{shim_JPF89,amoskov_Faraday94} and was 
also obtained in early computer
simulations~\cite{lai_JCP91,wittmer_JCP94}. Most recently,
refined theoretical models~\cite{biesheuvel_MM08} and 
computer simulations~\cite{he_MM07,merlitz_MM08,coluzza_PRL08} 
have delivered a considerable progress in terms of a quantitative 
understanding of polymer brushes at high grafting densities.

In our simulations, the polymers were created as a coarse-grained bead-spring model without
explicit twist or bending potential, i.e.\ the bonds were freely rotating and
freely jointed within the restrictions of excluded volume interactions.
The ``spring'' was a finite extensible nonlinear elastic 
(FENE) potential~\cite{kremer_JCP90}. The beads represent spherical
Kuhn monomers which interact via a shifted Lennard-Jones (LJ) potential
\begin{equation} \label{eq:10}
U(r) = 4\epsilon\left[ \left( \frac{d}{r} \right)^{12} -\left(
    \frac{d}{r}\right)^6 -\left( \frac{d}{r_c} \right)^{12} + \left(
    \frac{d}{r_c}\right)^6 \right]\,,
\end{equation}
where $d$ stands for the bead size and $\epsilon$ defines the potential
depth. The parameter $r_c$ is the cutoff distance: When cutting at
the potential minimum, $r_c = r_{\rm min} = 2^{1/6}d$, the attractive 
tail of the pair interaction is removed and the chain monomers
display an athermal behavior. A longer cutoff distance, however, retains the attraction, 
and now the pair interaction is exhibiting a temperature dependence 
which enters through the excluded volume interaction~\cite{rubinstein04}. 
For the monomer-monomer interaction, the potential was cut at $r_{\rm min}$,
but all interactions involving the modified end-groups were cut at 
$r_c = 2 r_{\rm min}$ to account for a temperature dependent solvent quality.

The simulations were carried out using the open source LAMMPS molecular
dynamics package~\cite{plimpton_CompPhys95}. In this paper, the
LJ system of units is used, with a monomer bead-size of $d = 1$,
mass $m = 1$, and a potential depth of $\epsilon = 1$, which 
defines both energy and temperature units (using a Boltzmann constant 
$k_B = 1$). The modified end-monomers
were of different sizes as will be discussed below in the context of the 
simulation results. The equation of motion of any non-grafted monomer 
was given by the Langevin equation, in which a friction term
was implemented as well as a random fluctuation term to create a
diffusive motion at well defined temperature and to account
for the implicit solvent. Further details about the simulation 
procedure are found in a previous publication~\cite{he_MM07}.

\begin{figure}[t]
\centerline{
\includegraphics[width=0.5\columnwidth]{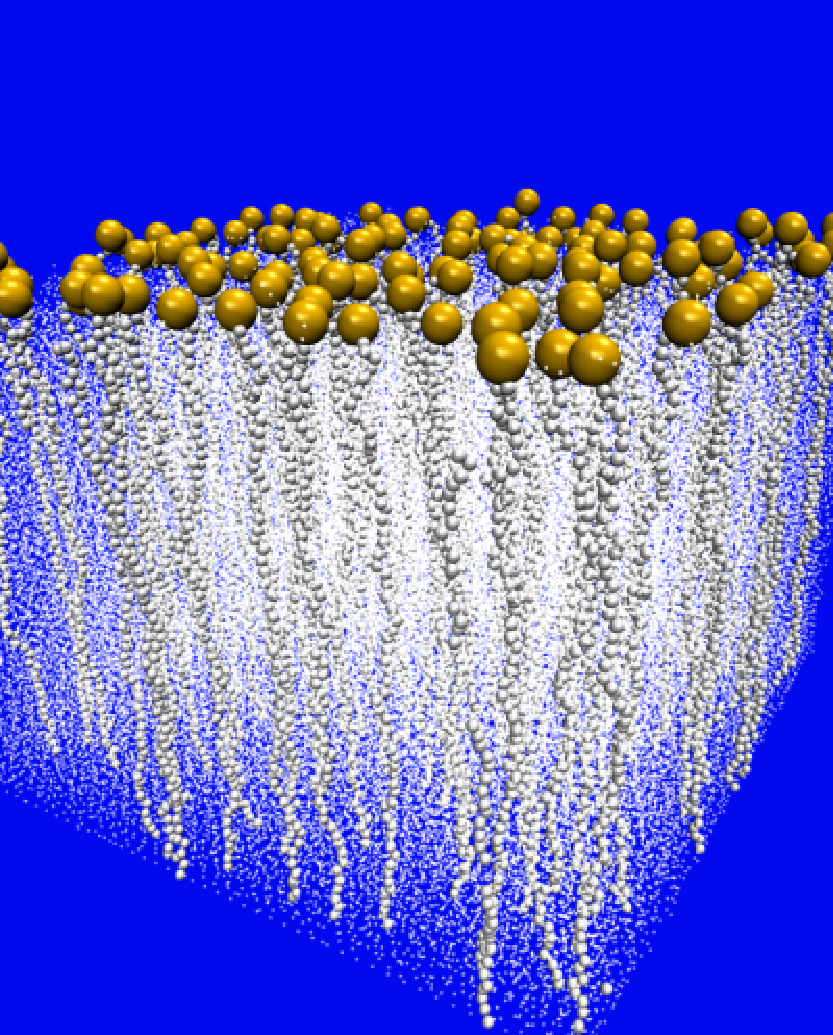}
\includegraphics[width=0.5\columnwidth]{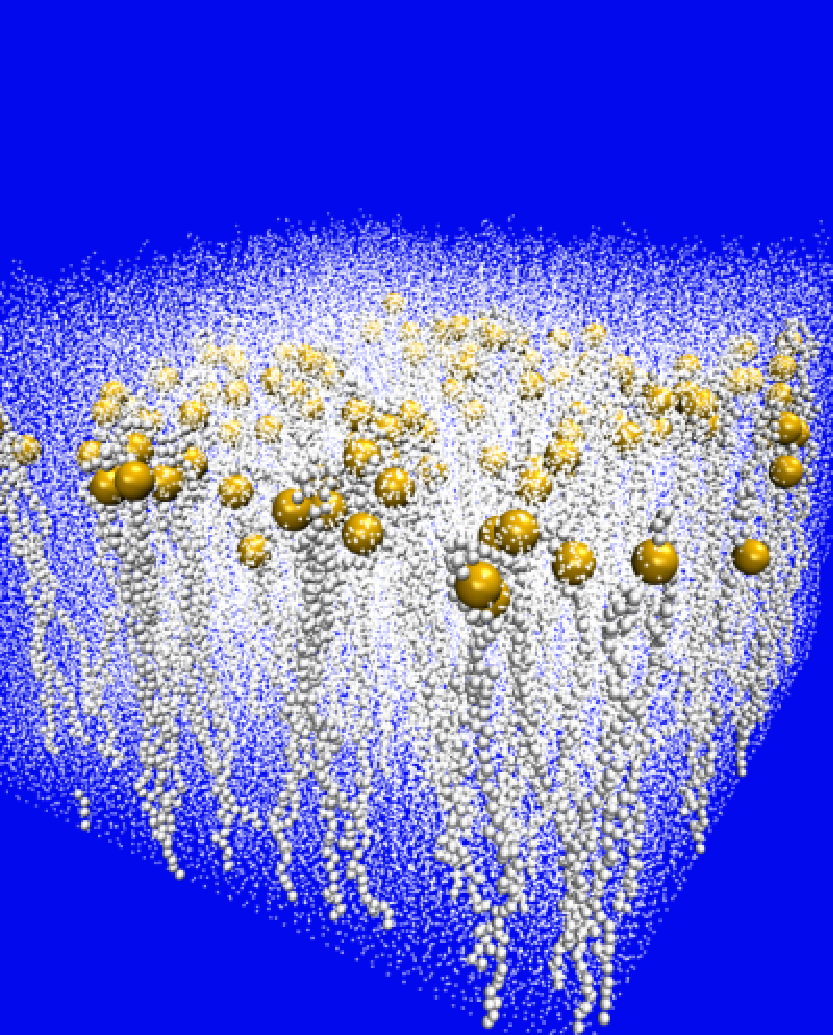}
}
\caption{Surface switching: In good solvent, the modified end-monomers
are ``swimming'' on top of the brush (left), in poor solvent, the
corresponding chains are retreating (right). The 95\% majority of unmodified 
chains are plotted transparent.
\label{fig:snapshot}}
\end{figure}

For our simulations, $24\times 24$ polymer chains, each of them with
$N = 64$ monomers, were grafted in a Cartesian pattern onto
a planar surface with a grafting density $\sigma = 0.46$ chains per
unit area, yielding a high density brush with chains
stretched to roughly 70\% of their contour length, a value which
has already been achieved in laboratory~\cite{devaux_MM05}.
In both horizontal directions, the system had periodic boundaries
to avoid any spurious finite size effects.
The brush was assumed to be inside an athermal solvent, 
but a minority of 5\% of the chains was chosen randomly and their
end-monomers replaced with beads of different chemical properties.
It was assumed that these end-groups were incompatible with the
solvent, so that they would not mix with that solvent at low 
temperature. The 95\% majority chains remained unmodified
and shall be denoted as ``normal'' chains.

Figure \ref{fig:snapshot} displays two snapshots, one of them taken at
high temperature (left), and another one at low temperature
(right). The modified end-monomers have got the diameter $d = 3$,
trice the size of the standard monomers. The observed switching of 
the end-groups from above to below the brush surface is fully 
reversible and can be repeated as often as desired. Of course,
whatever was facilitated via change of temperature in the simulation,
could equally well being achieved through a change of solvent in the
laboratory.

\begin{figure}[t]
\centerline{
\includegraphics[angle=270,width=1.0\columnwidth]{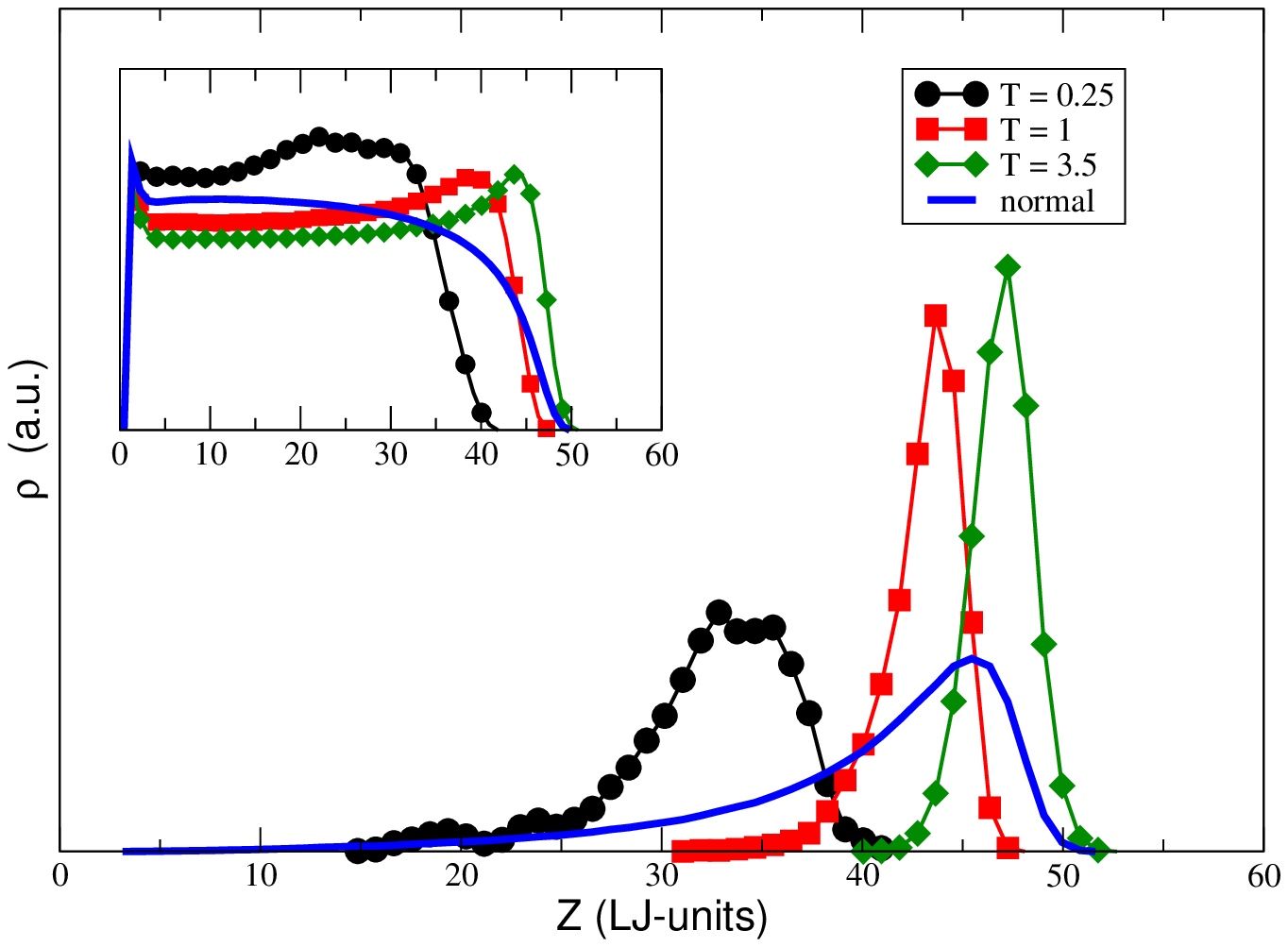}
}
\caption{Vertical density profiles: Distribution of the modified
end-groups of size $d = 3$ at different temperatures (the chains are
fixed to the substrate at $z = 0$). For comparison,
the normal (unmodified) chain ends (solid line). Inset: Density 
profiles of the entire chains. The vertical coordinate Z is given
in Lennard-Jones units, the density $\rho$ in arbitrary units.
\label{fig:profiles}}
\end{figure}

Figure \ref{fig:profiles} displays how the density distributions
of the end-groups vary with the system temperature. At 
$T = 0.25$ (circles), they remain well below the brush surface 
as defined with the normal, i.e.\ unmodified chain ends 
(solid line). Around
$T = 1$ (squares), the transition occurs, and at $T = 3.5$
the end-groups are well localized on top of the brush. The inset
shows how the density profiles of the entire chains are affected.
The normal chains are forming the box-like profile, characteristic
for high density brushes, being almost flat inside and 
dropping steeply near the surface. At low temperature, the modified 
chains are collapsing into the brush, at high temperature, they
become over-stretched and exhibit a density maximum at the brush 
surface.

\begin{figure}[t]
\centerline{
\includegraphics[angle=270,width=1.0\columnwidth]{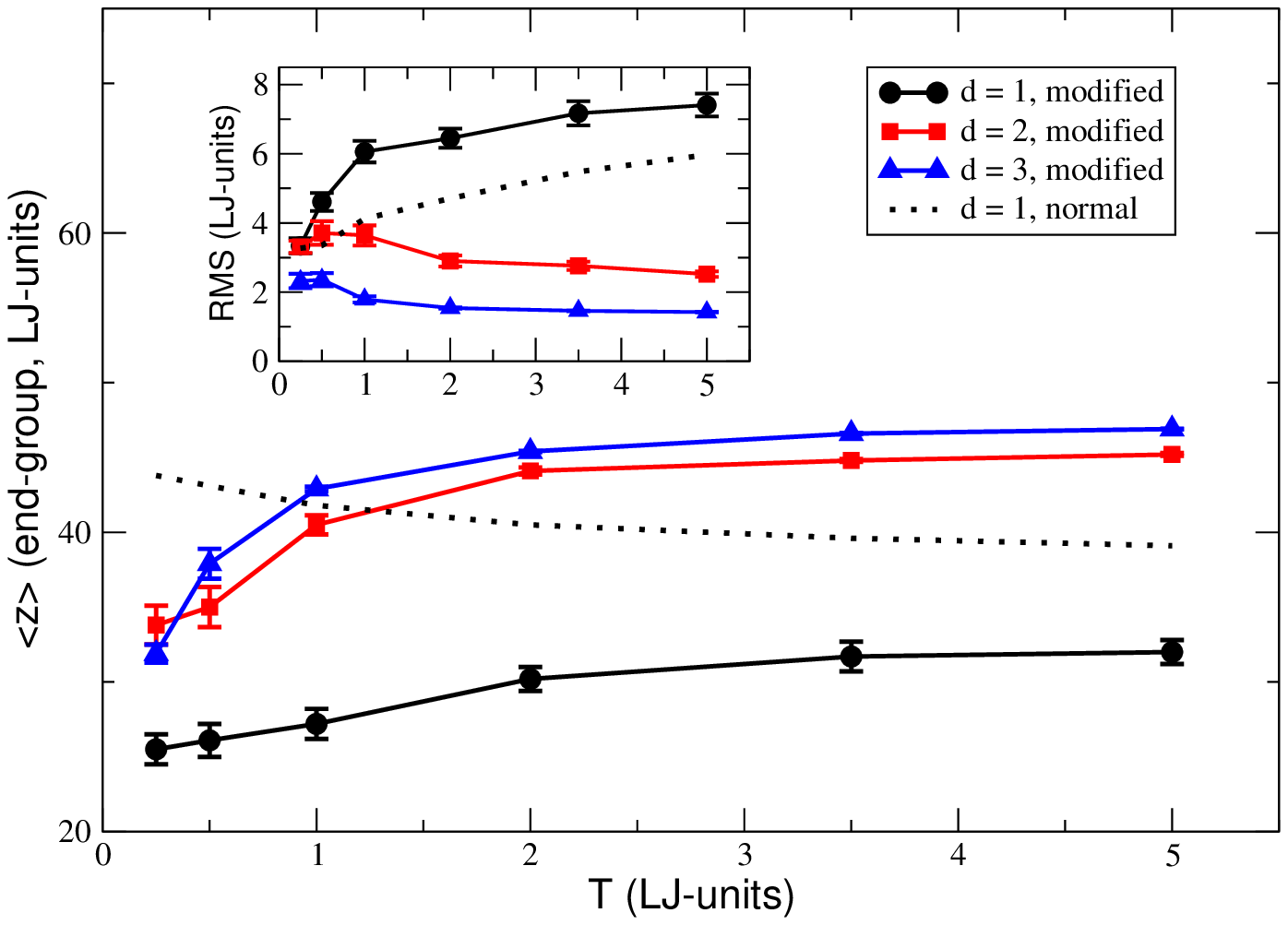}
}
\caption{Importance of end-group size: Average height of the
end-monomer as a function of temperature $T$ for different
end-group diameters. The dotted line is the brush surface, 
defined as the average position of the unmodified end-monomer. 
The grafting density is
$\sigma = 0.46$. Inset: The vertical rms fluctuations. Axes are
scaled in Lennard-Jones units.
\label{fig:sig04}}
\end{figure}

The system is asking 
for a couple of conditions to be satisfied in order to 
make such a brush functional. Figure \ref{fig:sig04}
displays the average heights and vertical rms-fluctuations
(inset) of the chain-ends as a function of temperature and
for three different end-group diameters. The dotted line 
corresponds to the unmodified ``normal'' chain-ends which
are athermal and have got the standard diameter $d = 1$.
It is clearly visible how the functional end-groups of size
$d > 1$ penetrate the surface at a temperature $T \approx 1$ and 
stay on top of the brush at any higher temperature (squares
and triangles). The 
end-group of size $d = 1$, however, does never switch (circles).
The vertical rms-fluctuations (inset) exhibit a decrease of
fluctuations for the inflated end-groups with increasing temperature.
This is an interesting observation, since it implies that these
groups stay pinned on top of the brush without significant
vertical fluctuations, an important condition to make the 
surface modification fully functional. There is no such
pinning effect with the modified group of size $d = 1$,
which continues to diffuse all through the brush at any
temperature (i.e.\ solvent quality).      

Why is the size of the end-group of the essence? The explanation
is based on the results found in recent studies on high-density
polymer brushes~\cite{merlitz_MM08,merlitz_MTS08}. The box-like
density profile of such a brush, as shown in Fig.\ \ref{fig:profiles}
(inset, solid line), exhibits a sharp density drop within a
narrow surface layer, and remains almost flat otherwise. Now
it is important to recall that any density gradient is generating
an osmotic pressure which is pushing the monomer towards the
direction of lowest density. Consequently, a high density 
brush contains a zone, localized just below the surface, in which
any monomer is exposed to a strong push upwards,
while the osmotic pressure deep inside the brush remains insignificant. 
This is fundamentally different from the situation with low-density
brushes, the vertical profiles of which are 
parabolic in good solvent~\cite{semenov_JETP85,milner_MM88}
and thus distributing the osmotic pressure all the way
through the brush and lacking any steep density gradient
near the surface. As was shown in~\cite{merlitz_MM08},
minority chains, when being slightly longer than the standard chain, or 
alternately end-modified with an inflated end-monomer,
were being pulled all the way through the osmotic pressure zone 
and exhibited a stretching energy of roughly twice the amount
measured at the normal (majority) chains.
To the contrast, chains which were just
a little shorter than average or which had got a reduced end-monomer 
size, tend to collapse into the brush with a diminishing probability to 
reach up to the surface. Intermediate states are strongly suppressed
because of the large amount of free energy required to stretch the
chain up to reach the osmotic pressure zone.
In this way, polydispersity effects
are amplified, creating a surface instability inside high density brushes.

\begin{figure}[t]
\centerline{
\includegraphics[width=1.0\columnwidth]{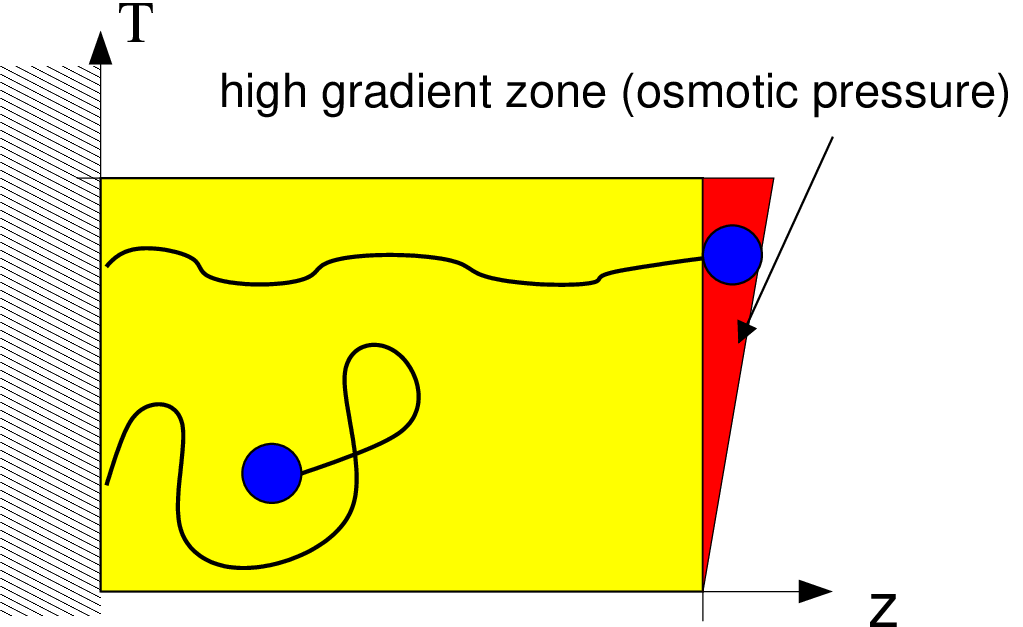}
}
\caption{Surface instability: Sketch of chain conformations with a modified
  end group in a high density brush. The box-like density profile
creates a narrow zone of osmotic pressure near the surface,
which, at high temperature, is pushing the 
end-group out of the brush (upper chain). 
At low temperature, the poor-solvent interaction with the
end-group is compensating the osmotic pressure and the chain is 
collapsing (lower chain).}
\label{fig:switchsketch}
\end{figure}

In the present system, the localization of the osmotic pressure zone
is exploited to facilitate an effectively switching surface.
At low temperature, the interaction of the poor solvent with the 
end-group generates an effective force which is pushing the chain-end 
below the surface. Alternately, one may discuss the effective size of that 
chain-end, which now represents a single attractive particle inside a melt
of athermal monomers. At low temperature, its excluded volume remains below
the monomer volume, and the modified chain therefore behaves the same way as the 
short minority chain discussed above: It stays in a 
collapsed state inside the brush (Fig.\ref{fig:switchsketch}, lower 
chain). With increasing temperature, its excluded
volume inside the melt is exceeding that of the athermal monomers and the chain,
now effectively being marginally longer than average, is penetrating the
osmotic pressure zone and stretched above
the brush surface (Fig.\ref{fig:switchsketch}, upper chain). 
If, however, the end-group is of equal size as the athermal monomers,
then its excluded volume cannot exceed their volume at any finite 
temperature, and the switch cannot occur (circles in Fig.\ \ref{fig:sig04}).

\begin{figure}[t]
\centerline{
\includegraphics[angle=270,width=1.0\columnwidth]{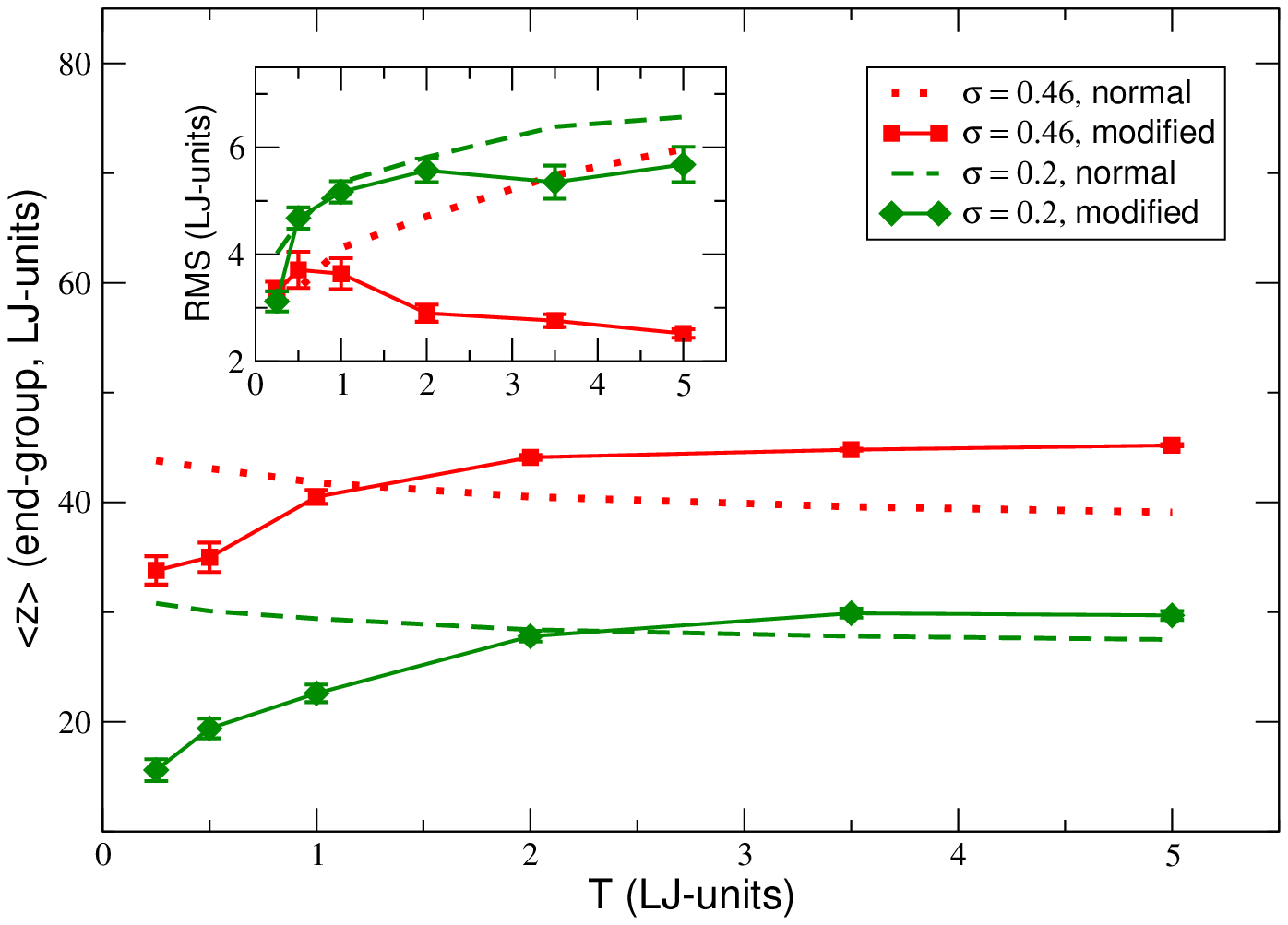}
}
\caption{Influence of grafting density $\sigma$: Average height of the
end-monomer as a function of temperature $T$ at different grafting
densities. The dotted and dashed lines correspond to unmodified
(normal) majority chains. The size of the modified
end-groups is $d = 2$. Inset: The vertical rms fluctuations.
Axes are scaled in Lennard-Jones units.
\label{fig:d2}}
\end{figure}

When considering the fact that the above discussed localization
of the osmotic pressure occurs particularly at high grafting densities,
then it is obvious that the observed switching of end-groups
should be less effective at moderate grafting densities. Figure
\ref{fig:d2} contains a comparison of two brushes, one of them
at high grafting density ($\sigma = 0.46$) and the second at
moderate density ($\sigma = 0.2$) which corresponds to a chain
stretch of roughly 45\% of the contour length. The switch does
still take place, though at a higher temperature $T \approx 2$
(diamonds) instead of  $T \approx 1$ with the high density 
brush (squares).
The rms-fluctuations (inset) however indicate that the end-groups
now exhibit strong vertical fluctuations and do not stay pinned
at the surface as they did at high grafting density. At moderate 
grafting densities, the vertical density gradient is much smoother and
extends over a longer distance below the brush surface, and hence 
the amplification of chain-end modifications and their effects
on chain statistics is significantly reduced~\cite{merlitz_MM08}.
Such a brush at $\sigma = 0.2$ could still be used to create
a functional surface through end-group modification, but in
that case the
switch of the surface properties would be less pronounced 
than with high density brushes, due to the enhanced 
vertical mobility of the end-groups.  

In summary, a homopolymer brush can be used to create a reversibly
switchable surface through end-group modification. In order to
achieve a highly efficient switch, a couple of conditions have to
be satisfied. First, the brush density should be high,
to generate a surface instability which amplifies the 
interactions of modified end-groups with the solvent. Then, the
modified ends should remain the minority because it requires 
the unmodified majority chains to maintain a steep density gradient 
near the surface. As another condition, the modified
end-groups have to be over-sized in order to push through the 
surface under good solvent (or high temperature) conditions and stay
put without major vertical fluctuations.
Finally, The present simulations were carried out under
ideal conditions in the
sense that a monodisperse environment of majority chains
was assumed, which is not perfectly achievable under laboratory
conditions. An increasing degree of polydispersity would lead to a
diffusion of the osmotic pressure zone and thereby reduce 
the amplification of solvent interactions with the 
end-groups, similar to the behavior found at reduced grafting density 
in Fig.\ \ref{fig:d2}.
Although a certain amount of polydispersity would 
be allowable without dramatically affecting its function,
further detailed studies are required
to analyze how exactly the efficiency of switching would depend on 
the degree of polydispersity in the system.     

Unlike traditional approaches to switchable surfaces via mixed
polymer brushes, the modification of end-groups would be technically 
rather simple and appears highly flexible, since a huge variety of
groups, including mixtures of them, could be attached to the chain ends
and allow for the creation of truly ``smart'' surfaces.

\begin{acknowledgments}
This work was partly supported by the National 
Science Foundation of China under Grant No. 10225420. G-L He
thanks J-U Sommer for the hospitality during a research
stay at the IPF Dresden.
\end{acknowledgments}

\bibliography{Brushes}

\end{document}